\documentclass[pra,twocolumn,superscriptaddress,showpacs,preprintnumbers,amsmath,amssymb,floatfix]{revtex4-2}
\usepackage{graphicx}% Include figure files
\usepackage{dcolumn}% Alig table columns on decimal point
\usepackage{bm}% bold math
\usepackage{enumerate}
\usepackage{mathdots}
\usepackage[table]{xcolor}
\begin{document}
\title{Floquet dynamics of Rabi model beyond the counterrotating hybridized rotating wave method}

\author{Yingying Han}
\email[Corresponding email: ]{hanyingying@sztu.edu.cn}
\affiliation{Shenzhen Key Laboratory of  Ultraintense Laser and Advanced Material Technology, Center for Intense Laser Application Technology, and College of Engineering Physics, Shenzhen Technology University, Shenzhen 518118, China}

\author{Shuanghao Zhang}
\affiliation{Shenzhen Key Laboratory of  Ultraintense Laser and Advanced Material Technology, Center for Intense Laser Application Technology, and College of Engineering Physics, Shenzhen Technology University, Shenzhen 518118, China}

\author{Meijuan Zhang}
\affiliation{Shenzhen Key Laboratory of  Ultraintense Laser and Advanced Material Technology, Center for Intense Laser Application Technology, and College of Engineering Physics, Shenzhen Technology University, Shenzhen 518118, China}
\affiliation{Institute of Theoretical Physics, Shanxi University, Taiyuan 030006, China}

\author{Q. Guan}
\affiliation{Department of Physics and Astronomy, Washington State University, Pullman, Washington 99164-2814, USA}

\author{Wenxian Zhang}
\affiliation{Key Laboratory of Artificial Micro- and Nano-structures of Ministry of Education, and School of Physics and Technology, Wuhan University, Wuhan, Hubei 430072, China}

\author{Weidong Li}
\affiliation{Shenzhen Key Laboratory of  Ultraintense Laser and Advanced Material Technology, Center for Intense Laser Application Technology, and College of Engineering Physics, Shenzhen Technology University, Shenzhen 518118, China}

\date{\today}

\begin{abstract}
Monochromatically driven two-level systems (i.e., Rabi models) are ubiquitous in various fields of physics. Though they have been exactly solved, the physical pictures in these exact solutions are not clear. Recently, approximate analytical solutions with neat physics have been obtained by using the counterrotating hybridized rotating wave (CHRW) method, which has been proven to be effective over a wider range of parameters than the previous analytical solutions. However, the CHRW depends on a parameter $\xi$, which has no solution in some regimes. Here we combine the double-unitary-transformation approach with the generalized Van Vleck nearly degenerate perturbation theory, and present approximate analytical results with clear physics for almost all parameter regimes, which agree well with the numerical solutions and the previous experimental results. Moreover, the dynamic frequencies of the Rabi model are regular, and the frequency with the highest Fourier amplitude changes from the Rabi frequency to $2n\omega$ with driving frequency $\omega$ and integer n, as the driving intensity increases from weak to deep-strong. In addition, we further explore the Floquet dynamics of the dissipative open Rabi model. Remarkably, the dissipations are tunable in the rotating frame, and the approximate analytical results obtained by our method are in good agreement with the numerical results in the strong driving regime. These results pave the way to quantum control using strong and deep-strong driving with applications in quantum technologies.

\end{abstract}

\maketitle

\section{Introduction}

Optical-control of materials based on Floquet engineering has been attracting great interest ranging from the realization of novel phenomena ~\cite{PhysRevLett.117.090402, SachaTime,Fenner2020Topology,Shevchenko20101,HanPRApplied,Han:22,Eckardt2017Atomic,PhysRevA.77.053601} to optical suppression of decoherence~\cite{PhysRevLett.114.190502, PhysRevB.75.201302, PhysRevB.77.125336}. The prototype of optically controlled quantum system is the Rabi model, describing a two-level system driven by a monochromatic field with amplitude $A$ and frequency $\omega$, which is common in different physical setups, ranging from quantum optics to condensed matter and quantum information~\cite{QO,Jaynes-Cummings}. For a weak and near-resonance driving, one usually invokes the rotating-wave approximation (RWA) and results show that the dynamics of the Rabi models are periodic oscillations with Rabi frequency~\cite{QO,Grifoni1998Driven}. In fact, the dynamics of the Rabi model are much richer than those under the RWA, such as Rabi oscillations' collapse and revival~\cite{PhysRevLett.44.1323}. To observe these and other novel phenomena in the lab, a strong (deep-strong) driving regime is required, that is, the driving amplitude $A$ has to be comparable or larger than the system's decoherence rates (transition frequency of the system)~\cite{PhysRevLett.105.263603,PhysRevB.96.014201,PhysRevLett.119.053203,Nori2009Quantum,RevModPhys.91.025005,PhysRevA.75.063414,
PhysRevLett.84.2112,PhysRevB.72.195410}. Beyond the weak driving regime, the RWA breaks down, and the frequency components of the dynamics are no longer a single Rabi frequency and become analytically difficult to predict.

An exact analytical solution to this Rabi model in terms of two known special functions has been presented in Ref.~\cite{PhysRevA.82.032117}, and an iterating approach for strong-coupling periodically driven two-level systems is presented in Ref.~\cite{PhysRevLett.98.013601}. Though they all show analytical results that agree well with the numerical results, it is difficult to analytically provide the frequency components of the system dynamics, which are demanded in experiments for the strongly driven system~\cite{PhysRevLett.115.133601}. Therefore, analytical solutions with clear physics are also needed. Counterrotating hybridized rotating wave (CHRW) approach is beyond the traditional RWA and remains the RWA form with a renormalized tunneling strength and a modified driving~\cite{PhysRevA.86.023831}. However, it is based on a unitary transformation with a parameter $\xi$, which has no solution in some situations. In this work, we present approximate analytical results of the Rabi model with clear physics for almost all parameter regimes by combining the double-unitary-transformation (DUT) approach with the generalized Van Vleck (GVV) nearly degenerate perturbation theory. Our analytical results agree well with the numerical solutions obtained by Floquet theory and the previous experimental results ~\cite{PhysRevLett.115.133601}. Moreover, we further investigate the effectiveness of the GVV method in the dissipative open Rabi model and find that it is still applicable in the strong driving regime.

The paper is organized as follows. The formalism of Floquet dynamics is presented in Sec.~\ref{sec:Floquet}. We explore the Floquet dynamics of the closed Rabi model in Sec.~\ref{sec:crm} and the dissipative open Rabi model in Sec.~\ref{sec:orm}. Conclusions are given in the last section of the article.

\section{Floquet dynamics}

\label{sec:Floquet}
In this work, we investigate the Rabi model in the Floquet picture: copy infinite two-level systems and then shift the energy separation between each copy by $\hbar\omega$ with Planck’s constant $\hbar$. Note that hereafter $\hbar=1$. This correspondence allows one to view Rabi oscillations and Floquet states from the simpler perspective of their time independent-problem equivalent~\cite{10.1119/10.0001897}. A periodic time-dependent Hamiltonian $H(t)=H(t+T)$ with period $T$ can be transformed to an equivalent time-independent infinite-dimensional Floquet matrix eigenvalue problem. The Hamiltonian $H(t)$ has Fourier components of $\omega$ with $\omega=2\pi/T$,
\begin{equation}
\label{eq:fhf}
  H(t) =\sum_{n}H^{[n]} \exp(-in\omega t),
\end{equation}
where the component $H^{[n]}$ is represented in the Floquet-state $|\alpha,n\rangle=|\alpha\rangle\bigotimes|n\rangle$ with the system index $\alpha$ (Note that in the generalized Floquet formalism $\alpha$ can be the N-level system index, but in this work, the Rabi model is restricted to N=2) and the Fourier index $n$ that runs from $-\infty$ to $\infty$~\cite{Chu2004Beyond}. According to the Floquet theory~\cite{Kohler2005Driven}, the elements of the infinite-dimensional Floquet matrix $H_{F}$ are defined by
\begin{equation}\label{eq:nfm}
 \langle \alpha, n|H_{F}|\beta, m\rangle=H_{\alpha \beta}^{[n-m]} + n\omega\delta_{\alpha\beta}\delta_{n m},
\end{equation}
with integers $n$ and $m$.
The Floquet matrix is then diagonalized,
\begin{equation}\label{eq:fme}
  H_{F}|\varepsilon_{\gamma l}\rangle=q_{\gamma l}|\varepsilon_{\gamma l}\rangle,
\end{equation}
where $q_{\gamma l}$ is the quasienergy eigenvalue and $|\varepsilon_{\gamma l}\rangle$ is the corresponding eigenvector. An initial state $|\Psi(0)\rangle$ can be written as
\begin{equation}
\label{eq:init}
  |\Psi(0)\rangle =\sum_{\gamma l}a_{\gamma l}|\varepsilon_{\gamma l}\rangle,
\end{equation}
with $a_{\gamma l}=\langle\varepsilon_{\gamma l}|\Psi(0)\rangle$. The wave function at time $t$ is
\begin{equation}
\label{eq:init}
  |\Psi(t)\rangle \!=\!\sum_{\gamma l}a_{\gamma l}e^{-i q_{\gamma l}t}|\varepsilon_{\gamma l}\rangle\!=\!\sum_{\gamma l}e^{-i q_{\gamma l}t|\varepsilon_{\gamma l}\rangle\langle\varepsilon_{\gamma l}|\Psi(0)\rangle}.
\end{equation}

For a given initial state $|\Psi(0)\rangle =|\alpha,0\rangle$, the time-averaged transition probability
from $|\alpha\rangle$ to $|\alpha'\rangle$ can be calculated:
\begin{eqnarray}\label{eq:thp}
  P_{\alpha\rightarrow \alpha'}(t)&=&|\sum_{n}\langle\alpha',n|\Psi(t)\rangle|^2 \nonumber\\
  &=&|\sum_{n}\sum_{\gamma l}e^{-i q_{\gamma l}t}\langle{\alpha',} n|\varepsilon_{\gamma l}\rangle\langle \varepsilon_{\gamma l}|\alpha,0\rangle|^{2}.
\end{eqnarray}
According to the Floquet theory~\cite{Shirley1965Solution}, the eigenvalues of the Floquet matrix exhibit translational symmetry, i.e., $q_{\gamma n}=q_{\gamma 0}+n\omega$. As a result, the probability to find the system in state $|\alpha'\rangle$ is expected to show oscillatory behavior with frequencies $n\omega$ and $\pm\Delta\varepsilon+n\omega$. Here, $\Delta\varepsilon=q_{\gamma n}-q_{\gamma' n}$. In the next section, we will focus on the analytical results of these frequencies.

\section{Closed Rabi model}
\label{sec:crm}
The Rabi model describes a two-level system, denoted by ground state $|0\rangle$ and excited state $|1\rangle$, driven by a harmonic driving with amplitude $A$ and frequency $\omega$,
\begin{equation}\label{eq:rabi}
H_{\text{Rabi}}(t)=\frac{\omega_0}{2}\sigma_z+\frac{A}{2}\cos(\omega t)\sigma_x,
\end{equation}
where $\sigma_{x,y,z}$ is the usual Pauli matrix and $\omega_0$ the transition frequency of the system. Recently, strongly driven quantum systems have attracted considerable attention~\cite{PhysRevX.7.041008,PhysRevA.75.063414}, and the traditional RWA is not valid in this regime. Thus, we will analytically investigate the Floquet dynamics of the Rabi model by using the CHRW~\cite{PhysRevA.91.053834} and GVV methods.

\subsection{ Counterrotating hybridized rotating wave }
\label{sec:chrw}
The essence of the CHRW method is a unitary transformation
\begin{equation}\label{eq:u1}
  U_1(t)=\text{exp}\left[-i\frac{A\xi}{2\omega}\sin(\omega t)\sigma_x\right],
\end{equation}
with parameter $\xi\in[0,1]$ to be determined later~\cite{PhysRevA.86.023831}. The Hamiltonian in Eq.~(\ref{eq:rabi}) after the transformation is
\begin{eqnarray}\label{eq:u1ha}
  H_1(t)&=&U_1^\dag(t)H_{\text{Rabi}}(t)U_1(t)-i U_1^\dag(t)\frac{\partial U_1(t)}{\partial t} \nonumber \\
  &=&\frac{\omega_0}{2}
  \left\{\cos\left[\frac{A\xi}{\omega}\sin(\omega t)\right]\sigma_z+\sin\left[\frac{A\xi}{\omega} \sin(\omega t)\right]\sigma_y\right\} \nonumber\\
  &+&\frac{A}{2}(1-\xi)\cos(\omega t)\sigma_x.
\end{eqnarray}
Note that after the transformation $U_1(t)$, the basis states of the Hamiltonian in Eq.~(\ref{eq:u1ha}) are
\begin{eqnarray}\label{eq:u1basis}
  |s_{1}(t)\rangle&\!=\!&\cos\!\left[\frac{A\xi}{2\omega}\sin(\omega t)\right]\!|1\rangle\!+\!i\sin\!\left[\frac{A\xi}{2\omega}\sin(\omega t)\right]\!|0\rangle,\nonumber \\
  |s_{0}(t)\rangle&\!=\!&i\sin\!\left[\frac{A\xi}{2\omega}\sin(\omega t)\right]\!|1\rangle\!+\!\cos\!\left[\frac{A\xi}{2\omega}\sin(\omega t)\right]\!|0\rangle. \nonumber\\
\end{eqnarray}
Using $\text{exp}[i A\xi/\omega\sin(\omega t)]=\sum_{n=-\infty}^{\infty}J_n(A\xi/\omega)\text{exp}(in\omega t)$ with $n$th-order Bessel function of the first kind $J_n(\cdot)$, Eq.~(\ref{eq:u1ha}) can be divided into three parts $H_1(t)=H_1'+H_1''(t)+H_1'''(t)$, where
\begin{eqnarray}
  H_1'&=&\frac{\omega_0}{2}J_0\left(\frac{A\xi}{\omega}\right)\sigma_z,\\
  H_1''(t)&=&\frac{A}{2}(1-\xi)\cos(\omega t)\sigma_x+\omega_0J_1\left(\frac{A\xi}{\omega}\right){\sin(\omega t)}\sigma_y,\nonumber\\\label{eq:u1h123}
\end{eqnarray}
and $H_1'''(t)=H_1(t)-H_1'-H_1''(t)$ includes all higher-order harmonic terms $\sin(n\omega t)$ and $\cos(n\omega t)$ with $n\geq2$, which are ignored in the CHRW method. Therefore, $H_1(t)\simeq H_1'+H_1''(t)$ and can be rewritten as
\begin{equation}\label{eq:u1h12}
  H_1(t)=\frac{\omega_0}{2}J_0\left(\frac{A\xi}{\omega}\right)\sigma_z+\frac{\tilde{A}}{4}(e^{-i\omega t}\sigma_{+}+e^{i\omega t}\sigma_{-}),
\end{equation}
where
\begin{equation}\label{eq:chrw1}
\frac{A}{2}(1-\xi)=\omega_0J_1\left(\frac{A\xi}{\omega}\right)\equiv\frac{\tilde{A}}{4},
\end{equation}
with $\xi\in[0,1]$. Clearly, Eq.~(\ref{eq:u1h12}) possesses a RWA-like form with a renormalized transition frequency $J_0(A\xi/\omega)\omega_0$ and a renormalized driving strength $\tilde{A}$.

For an initial state $|0\rangle$, the population of excited state $P_1(t)$ can be easily given as (see Appendix~\ref{app:chrw})
\begin{eqnarray}\label{eq:chrwp2}
P_1(t)&=&C_0+C_1\cos(\tilde{\Omega} t)+\sum_{n=1}^{\infty}[C_{2_n}\cos(2n\omega t+\tilde{\Omega} t)\nonumber \\
&+&C_{3_n}\cos(2n\omega t-\tilde{\Omega} t)+C_{4_n}\cos(2 n\omega t)],
\end{eqnarray}
with effective Rabi frequency
\begin{equation}\label{eq:effrabi}
\tilde{\Omega}=\sqrt{\tilde{\Delta}^2+\tilde{A}^2/4},
\end{equation}
and renormalized detuning $\tilde{\Delta}=J_0(A\xi/\omega)\omega_0-\omega$. The coefficients $C_0$, $C_1$, $C_{2_n}$, $C_{3_n}$ and $C_{4_n}$ with $n=1, 2, 3, \cdots, \infty$ in Eq.~(\ref{eq:chrwp2}) are the Fourier amplitudes of the cosine functions with frequencies $0$, $\tilde{\Omega}$, $2n\omega+\tilde{\Omega}$, $2n\omega-\tilde{\Omega}$ and $2n\omega$ respectively, and they are shown in Appendix~\ref{app:chrw}. Form Eq.~(\ref{eq:chrwp2}), we find that the dynamic frequency components of the Rabi model are infinite, and they exhibit translational symmetry with a translation of $2n\omega$, which is different from the general $n\omega$ discussed in Sec.~\ref{sec:Floquet}. Although the even photon number (i.e., $2n\omega$), has been discussed before ~\cite{Shirley1965Solution,PhysRevA.1.803,PhysRevLett.25.1149,2020Pulse}, here we present the analytical results of all the dynamic frequencies and corresponding amplitudes, which, to our knowledge, have not been presented directly before. The reason for the even photon number is that the periodic-driven signal in the Rabi model (the cosine signal) has additional symmetry~\cite{PhysRevLett.115.133601}. 

\begin{figure}[thp]
\includegraphics[width=3.2in]{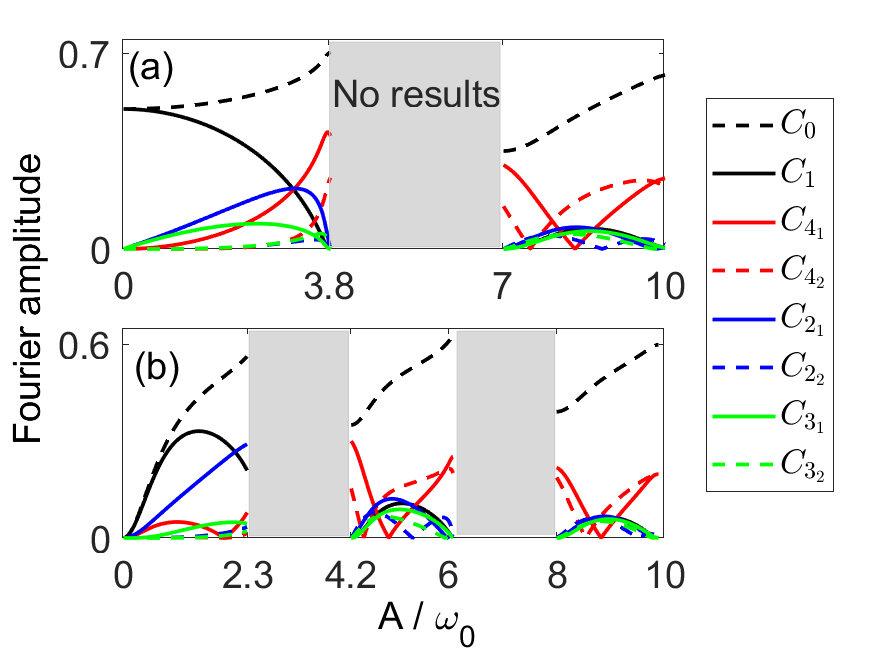}
\caption{\label{fig:chrw2}Coefficients $C_0$, $C_1$, $C_{4_1}$, $C_{4_2}$, $C_{2_1}$, $C_{2_2}$, $C_{3_1}$ and $C_{3_2}$ in Eqs.~(\ref{eq:chrwp2}) as a function of $A$ with $\omega/\omega_0=1$ (a) and $\omega/\omega_0=0.6$ (b), obtained from Eq.~(\ref{eq:c0})-(\ref{eq:c4}). The gray areas denote no results because there are no reasonable $\xi$ based on Eq.~(\ref{eq:chrw1}), i.e., no solution for $\xi\in[0,1]$ satisfies $\frac{A}{2}(1-\xi)=\omega_0J_1\left(\frac{A\xi}{\omega}\right)$.}
\end{figure}

\begin{figure}[thp]
\includegraphics[width=3.2in]{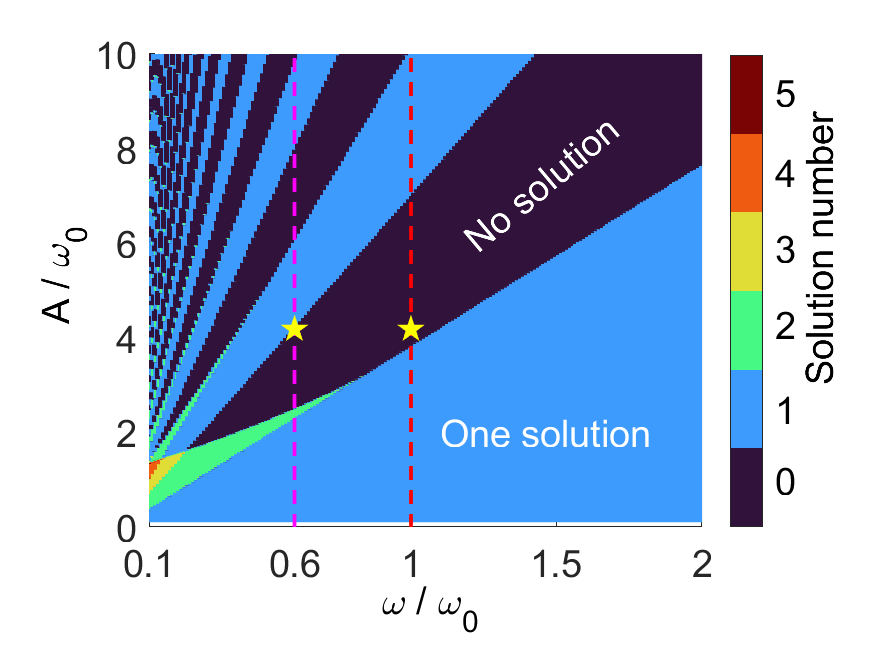}
\caption{\label{fig:solution} The number of solutions to $\xi\in[0,1]$ that satisfy equation $\frac{A}{2}(1-\xi)=\omega_0J_1\left(\frac{A\xi}{\omega}\right)$.} The magenta and red dashed lines are $\omega/\omega_0=0.6$ and $\omega/\omega_0=1$, respectively. The yellow stars on the magenta and red dashed lines represent the maximum driving strength (i.e., $A/\omega_0=4.18$) that were realized in experiments~\cite{PhysRevLett.115.133601}.
\end{figure}

The effectiveness of CHRW has been demonstrated~\cite{PhysRevA.86.023831}, and here we directly use this method to analyze the major frequencies of system dynamics at different driving intensities, which have not been studied previously. Figure~\ref{fig:chrw2} shows the variation of Fourier amplitude [i.e., $C_0$, $C_1$, $C_{4_1}$, $C_{4_2}$, $C_{2_1}$, $C_{2_2}$, $C_{3_1}$ and $C_{3_2}$ in Eq.~(\ref{eq:chrwp2})] with the driving intensity $A$. For a resonant situation in Fig.~\ref{fig:chrw2}(a), when $A/\omega_0\rightarrow 0$, the Floquet dynamic is a constant of 0.5 plus a Rabi oscillation with an amplitude of 0.5 and a frequency of $\tilde{\Omega}$. As $A$ increases, the Fourier amplitude of the Rabi oscillation with frequency $\tilde{\Omega}$ decreases, while the Fourier amplitudes of the high-frequency oscillation terms (i.e., $2n\omega\pm\tilde{\Omega}$ and $2n\omega$) increase and gradually exceed the former, indicating that the RWA without high-frequency oscillation terms is no longer valid. In addition, in most intervals of the deep-strong driving regime (i.e., $A/\omega_0\gg1$), the Fourier amplitudes of $2n\omega$ terms are larger than those of other terms and become the main oscillation terms of the Floquet dynamics. Moreover, we show the variation of the Fourier amplitudes with $A$ for the off-resonance situation in Fig.~\ref{fig:chrw2} (b). When $A/\omega\rightarrow 0$, the transition probability $P_1(t)\rightarrow 0$, which is reasonable for non-resonant weak coupling. The amplitude of the Rabi frequency $C_1$ increases with the increase of $A$, reaching its maximum at about $A/\omega_0=1$, and then gradually decreases. Similarly, $2n\omega$ terms become the main oscillation terms of the Floquet dynamics in the most intervals of deep-strong driving regime. Note that Eq.~(\ref{eq:chrwp2}) has no solution with some parameters in the gray areas as shown in Fig.~\ref{fig:chrw2}, which we will discuss in the next paragraph.

We present the valid conditions of the CHRW in this paragraph. From Eq.~(\ref{eq:chrw1}), we see that the solution of $\xi$ is the key of the CHRW. Figure~\ref{fig:solution} shows the number of solutions for the parameter $\xi\in[0,1]$ that satisfy equation $A(1-\xi)/2=\omega_0J_1\left(A\xi/\omega\right)$. We find that with parameters $\omega/\omega_0=0.27~(0.15), A/\omega_0=1.41~(1.35)$, there are two (three) solutions, i.e., $\xi=0.253,0.6043$ ($\xi=0.1691,0.2837,0.8256$), and so on. In these regions where $\xi$ has no unique solution, the CHRW method is no longer applicable. Also in some parameter ranges, no $\xi$ belongs to $[0,1]$  that satisfies $A(1-\xi)/2=\omega_0J_1\left(A\xi/\omega\right)$, such as $3.84\leq A/\omega_0\leq7.01$ with $\omega/\omega_0=1$ (shown by the red dashed line) and $2.30\leq A/\omega_0\leq4.20$ with $\omega/\omega_0=0.6$ (shown by the magenta dashed line). However, the maximum driving intensity experimentally studied reached $A/\omega_0=4.18$, as the yellow stars shown in Fig.~\ref{fig:solution}. Therefore, analytical results with clear physics suitable for the whole driving-strength range need to be solved urgently.

\subsection{Double-unitary-transformation }\label{sec:dut}
In this section, we use the DUT approach to investigate the Floquet dynamics of the Rabi model in Eq.~(\ref{eq:rabi}). DUT is a powerful method for dealing with transverse (off-diagonal) and periodically driven quantum systems (such as the Rabi model), which makes it possible to obtain analytical results by using perturbation theory. The first unitary transformation is $U_2=\text{exp}(-i\pi\sigma_y/4)$, and the Hamiltonian in Eq.~(\ref{eq:rabi}) after this transformation is
\begin{equation}\label{eq:dut1}
H_2(t)=\frac{A}{2}\cos(\omega t)\sigma_z-\frac{\omega_0}{2}\sigma_x,
\end{equation}
with basis $|\pm\rangle=(|1\rangle\pm|0\rangle)/\sqrt{2}$.
The second unitary transformation is
\begin{equation} \label{eq:dut2}
U_3(t)=\text{exp}\left[-i\frac{A}{2\omega}\sin(\omega t)\sigma_z\right],
\end{equation}
and the Hamiltonian in Eq.~(\ref{eq:dut1}) after this transformation becomes
\begin{eqnarray}\label{eq:dut3}
H_3(t)&=&\frac{-\omega_0}{2}\sum_{n=-\infty}^{\infty}J_n e^{-in\omega t}[(-1)^n \sigma_{+}+\sigma_{-}],
\end{eqnarray}
where $J_n(A/\omega)$ abbreviate as $J_n$ and $\sigma_{\pm}=(\sigma_x\pm i \sigma_y)/2$.
Note that the basis  states of Eq.~(\ref{eq:dut3}) are
\begin{eqnarray}\label{eq:dut4}
|s'_+(t)\rangle=\frac{\text{exp}[\frac{iA\sin(\omega t)}{2\omega}]|1\rangle+\text{exp}[\frac{-iA\sin(\omega t)}{2\omega}]|0\rangle}{\sqrt{2}},\\
|s'_-(t)\rangle=\frac{\text{exp}[\frac{iA\sin(\omega t)}{2\omega}]|1\rangle-\text{exp}[\frac{-iA\sin(\omega t)}{2\omega}]|0\rangle}{\sqrt{2}}.
\end{eqnarray}
According to Eq.~(\ref{eq:fhf}), the Floquet matrix block of Eq.~(\ref{eq:dut3}) is
\begin{eqnarray}
\label{eq:trans}
 H^{[n]} &=& \frac{-\omega_0}{2}J_n [(-1)^n \sigma_{+}+\sigma_{-}].
\end{eqnarray}
Combined Eq.~(\ref{eq:trans}) with Eq.~(\ref{eq:nfm}), the Floquet matrix of Eq.~(\ref{eq:dut3}) is
\begin{widetext}
\begin{equation}
\label{eq:dut5}
\renewcommand\arraystretch{1.7}
H_F=\left(
\begin{array}{c c c|c c|c c|c c|c c c}
\ddots&&&&&&&&&&&\iddots\\
 \hline
   &-2\omega & \frac{-\omega_0}{2}J_0&0& \frac{-\omega_0}{2}J_1 &0&\frac{-\omega_0}{2}J_2&0&\frac{-\omega_0}{2}J_3&0&\frac{-\omega_0}{2}J_4&\\
   &\frac{-\omega_0}{2}J_0&-2\omega&\frac{\omega_0}{2}J_1&0&\frac{-\omega_0}{2}J_2&0&\frac{\omega_0}{2}J_3&0&\frac{-\omega_0}{2}J_4&0&\\
   \hline
  &0&\frac{\omega_0}{2}J_1&-\omega &\frac{-\omega_0}{2}J_0&0&\frac{-\omega_0}{2}J_1&0&\frac{-\omega_0}{2}J_2&0&\frac{-\omega_0}{2}J_3&\\
  &\frac{-\omega_0}{2}J_1&0&\frac{-\omega_0}{2}J_0&-\omega&\frac{\omega_0}{2}J_1&0&\frac{-\omega_0}{2}J_2&0&\frac{\omega_0}{2}J_3&0&\\
  \hline
  &0&\frac{-\omega_0}{2}J_2&0 & \frac{\omega_0}{2}J_1& 0&\frac{-\omega_0}{2}J_0& 0&\frac{-\omega_0}{2}J_1&0&\frac{-\omega_0}{2}J_2&\\
  &\frac{-\omega_0}{2}J_2&0&\frac{-\omega_0}{2}J_1&0&\frac{-\omega_0}{2}J_0&0&\frac{\omega_0}{2}J_1&0&\frac{-\omega_0}{2}J_2&0&\\
  \hline
  &0&\frac{\omega_0}{2}J_3&0&\frac{-\omega_0}{2}J_2&0&\frac{\omega_0}{2}J_1 & \omega &\frac{-\omega_0}{2}J_0&0&\frac{-\omega_0}{2}J_1&\\
  &\frac{-\omega_0}{2}J_3&0&\frac{-\omega_0}{2}J_2&0&\frac{-\omega_0}{2}J_1&0&\frac{-\omega_0}{2}J_0&\omega&\frac{\omega_0}{2}J_1&0&\\
  \hline
  &0&\frac{-\omega_0}{2}J_4&0&\frac{\omega_0}{2}J_3&0&\frac{-\omega_0}{2}J_2&0&\frac{\omega_0}{2}J_1&2\omega&\frac{-\omega_0}{2}J_0&\\
  &\frac{-\omega_0}{2}J_4&0&\frac{-\omega_0}{2}J_3&0&\frac{-\omega_0}{2}J_2&0&\frac{-\omega_0}{2}J_1&0&\frac{-\omega_0}{2}J_0&2\omega&\\
   \hline
\iddots&&&&&&&&&&&\ddots
\end{array}
\right)
\renewcommand\arraystretch{1.7}
\begin{array}{c c}
\leftarrow&|s'_-,-2\rangle\\
\leftarrow&|s'_+,-2\rangle\\
\leftarrow&|s'_-,-1\rangle\\
\leftarrow&|s'_+,-1\rangle\\
\leftarrow&|s'_-,0\rangle\\
\leftarrow&|s'_+,0\rangle\\
\leftarrow&|s'_-,+1\rangle\\
\leftarrow&|s'_+,+1\rangle\\
\leftarrow&|s'_-,+2\rangle\\
\leftarrow&|s'_+,+2\rangle\\
\end{array}
\end{equation}

where $J_k=J_k(A/\omega)$. Then we perform a basis transformation with
\begin{equation}
\label{eq:dut6}
\renewcommand\arraystretch{1.2}
\setlength{\arraycolsep}{3pt}
S=\left(
\begin{array}{c c c|c c|c c|c c|c c c}
\ddots&&&&&&&&&&&\iddots\\
 \hline
   &1 & 1&0& 0 &0&0&0&0&0&0&\\
   &1&-1&0&0&0&0&0&0&0&0\\
   \hline
  &0&0&1&1&0&0&0&0&0&0\\
  &0&0&1&-1&0&0&0&0&0&0\\
  \hline
  &0&0&0 & 0& 1&1& 0&0&0&0&\\
  &0&0&0&0&1&-1&0&0&0&0&\\
  \hline
  &0&0&0&0&0&0&1&1&0&0&\\
  &0&0&0&0&0&0&1&-1&0&0&\\
  \hline
  &0&0&0&0&0&0&0&0&1&1&\\
  &0&0&0&0&0&0&0&0&1&-1&\\
   \hline
\iddots&&&&&&&&&&&\ddots
\end{array}
\right),
\end{equation}
and Eq.~(\ref{eq:dut5}) becomes
%\begin{widetext}
\begin{equation}
\label{eq:dut7}
\renewcommand\arraystretch{1.6}
\setlength{\arraycolsep}{0.8pt}
H'_F=
\left(
\begin{array}{c c c|c c|c c|c c|c c c}
\ddots&&&&&&&&&&&\iddots\\
 \hline
   &-2\omega-\frac{J_0\omega_0}{2} & 0&0& \frac{\omega_0J_1}{2} &\frac{-\omega_0J_2}{2} &0&0&\frac{\omega_0J_3}{2}&\frac{-\omega_0J_4}{2} &0&\\
   &0&-2\omega+\frac{J_0\omega_0}{2}&\frac{-\omega_0J_1}{2}&0&0&\frac{\omega_0J_2}{2}&\frac{-\omega_0J_3}{2}&0&0&\frac{\omega_0J_4}{2}&\\
   \hline
  &0&\frac{-\omega_0J_1}{2}&-\omega-\frac{J_0\omega_0}{2}&0&0&\frac{\omega_0J_1}{2}&\frac{-\omega_0J_2}{2}&0&0&\frac{\omega_0J_3}{2}&\\
  &\frac{\omega_0J_1}{2}&0&0&-\omega+\frac{J_0\omega_0}{2}&\frac{-\omega_0J_1}{2}&0&0&\frac{\omega_0J_2}{2}&\frac{-\omega_0J_3}{2}&0&\\
  \hline
  &\frac{-\omega_0J_2}{2}&0&0 &\frac{-\omega_0J_1}{2}&\frac{-\omega_0J_0}{2}&0& 0&\frac{\omega_0J_1}{2}&\frac{-\omega_0J_2}{2}&0\\
  &0&\frac{\omega_0J_2}{2}&\frac{\omega_0J_1}{2}&0&0&\cellcolor{blue!20}\frac{\omega_0J_0}{2}&\cellcolor{blue!20}\frac{-\omega_0J_1}{2}&0&0&\frac{\omega_0J_2}{2}&\\
  \hline
  &0&\frac{-\omega_0J_3}{2}&\frac{-\omega_0J_2}{2}&0&0&\cellcolor{blue!20}\frac{-\omega_0J_1}{2}&\cellcolor{blue!20}\omega-\frac{\omega_0J_0}{2} &0&0&\frac{\omega_0J_1}{2}&\\
  &\frac{\omega_0J_3}{2}&0&0&\frac{\omega_0J_2}{2}&\frac{\omega_0J_1}{2}&0&0&\omega+\frac{J_0\omega_0}{2}&\frac{-\omega_0J_1}{2}&0&\\
  \hline
  &\frac{-\omega_0J_4}{2}&0&0&\frac{-\omega_0J_3}{2}&\frac{-\omega_0J_2}{2}&0&0&\frac{-\omega_0J_1}{2}&2\omega-\frac{J_0\omega_0}{2}&0&\\
  &0&\frac{\omega_0J_4}{2}&\frac{\omega_0J_3}{2}&0&0&\frac{\omega_0J_2}{2}&\frac{\omega_0J_1}{2}&0&0&2\omega+\frac{J_0\omega_0}{2}&\nonumber \\
  \hline
\iddots&&&&&&&&&&&\ddots
\end{array}
\right).
\end{equation}
\begin{equation}
\renewcommand\arraystretch{1.6}
\setlength{\arraycolsep}{4.5pt}
\begin{array}{cccc c cccc cc c c c cc c c c c c}
&&&&&\uparrow&\uparrow&&&\uparrow&\uparrow&&\uparrow&\uparrow&\uparrow&\uparrow&&&\uparrow&\uparrow&\\
&&&&&|s'_0,-2\rangle&|s'_1,-2\rangle&&&|s'_0,-1\rangle&|s'_1,-1\rangle&&|s'_0,0\rangle&|s'_1,0\rangle&|s'_0,1\rangle&|s'_1,1\rangle&&&|s'_0,2\rangle
&|s'_1,2\rangle&
\end{array}
\end{equation}
\end{widetext}
with Floquet basis 
\begin{eqnarray}\label{eq:dut8}
  |s'_{1}(t)\rangle\!&=&\!\cos\left[\frac{A}{2\omega}\sin(\omega t)\right]\!|1\rangle\!+\!i\sin\left[\frac{A}{2\omega}\sin(\omega t)\right]|0\rangle\nonumber \\
  |s'_{0}(t)\rangle\!&=&\!i\sin\left[\frac{A}{2\omega}\sin(\omega t)\right]\!|1\rangle\!+\!\cos\left[\frac{A}{2\omega}\sin(\omega t)\right]|0\rangle .\nonumber \\
\end{eqnarray}
%One can see that $|s'_{1,0}(t)\rangle$ equal to $|s_{1,0}(t)\rangle$ with $\xi=1$.

From the matrix structure of $H'_F$ in Eq.~(\ref{eq:dut7}), one sees that $|s'_1,0\rangle$ couples to $|s'_0,1\rangle$ via an off-diagonal term of $-\omega_0J_1/2$, and this block matrix is shown in light blue. By tuning the frequency of the driving field (i.e., $\omega$), the Floquet state $|s'_1,0\rangle$ can become nearly degenerate with $|s'_0,1\rangle$, namely, $\omega\approx J_0\omega_0$. Following the standard GVV nearly degenerate perturbation theory and to the second order~\cite{PhysRevA.101.022108,PhysRevA.79.032301}, Eq.~(\ref{eq:dut7}) is reduced to a $2\times2$ matrix by including all other off-resonant coupling channels as perturbation terms (see Appendix~\ref{app:app1}):
\begin{eqnarray}\label{eq:gvv}
\renewcommand\arraystretch{1.4}
 H_{\rm GVV}=
 \left(
 \begin{array}{c c }
 \frac{J_0\omega_0}{2}+\delta_{1'} &\frac{-J_1\omega_0}{2}+\delta_{1'0'}\vspace{2mm}\\
 \frac{-J_1\omega_0}{2}+\delta_{0'1'}&\omega-\frac{J_0\omega_0}{2}+\delta_{0'}
 \end{array}
 \right),
\end{eqnarray}
where
\begin{eqnarray}\label{eq:delta1}
\delta_{1'}&=&\sum_{\begin{subarray}{c} k=-\infty \\ k\neq 0\end{subarray}}^{\infty}\left[\frac{(J_{2k+1}\omega_0/2)^2}{J_0\omega_0-(2k+1)\omega}+\frac{(J_{2k}\omega_0/2)^2}{-2k\omega}\right],\\
\delta_{0'}&\!=\!&\sum_{\begin{subarray}{c} k=-\infty \\ k\neq 0\end{subarray}}^{\infty}\left[\frac{-(J_{2k-1}\omega_0/2)^2}{J_0\omega_0+(2k-1)\omega}\!+\!\frac{(J_{2k}\omega_0/2)^2}{-2k\omega}\right],\\
\delta_{1'0'}&\!=\!&\sum_{\begin{subarray}{c} k=-\infty \\ k\neq 0\end{subarray}}^{\infty}\left[
\frac{-J_{2k-1}J_{2k}\omega_0^2/4}{J_0\omega_0\!+\!(2k-1)\omega}+\frac{J_{2k+1}J_{2k}\omega_0^2}{-8k\omega}\right],\\
\delta_{0'1'}&\!=\!&\sum_{\begin{subarray}{c} k=-\infty \\ k\neq 0\end{subarray}}^{\infty}\left[
\frac{J_{2k+1}J_{2k}\omega_0^2/4}{J_0\omega_0-(2k+1)\omega}\!+\!\frac{J_{2k-1}J_{2k}\omega_0^2}{-8k\omega}\right],\label{eq:delta4}
\end{eqnarray}
which result from the higher-order harmonic terms. Therefore, here our approach includes the higher-order harmonic terms that are ignored in CHRW. In Appendix~\ref{app:app1}, we numerically prove that $\delta_{0'}=-\delta_{1'}$, $\delta_{1'0'}=\delta_{0'1'}$ within the parameter range of our research and also demonstrate that the second-order perturbation results are more suitable for the exact numerical results than the first-order perturbation results.

\begin{figure}[thp]
\includegraphics[width=3.2in]{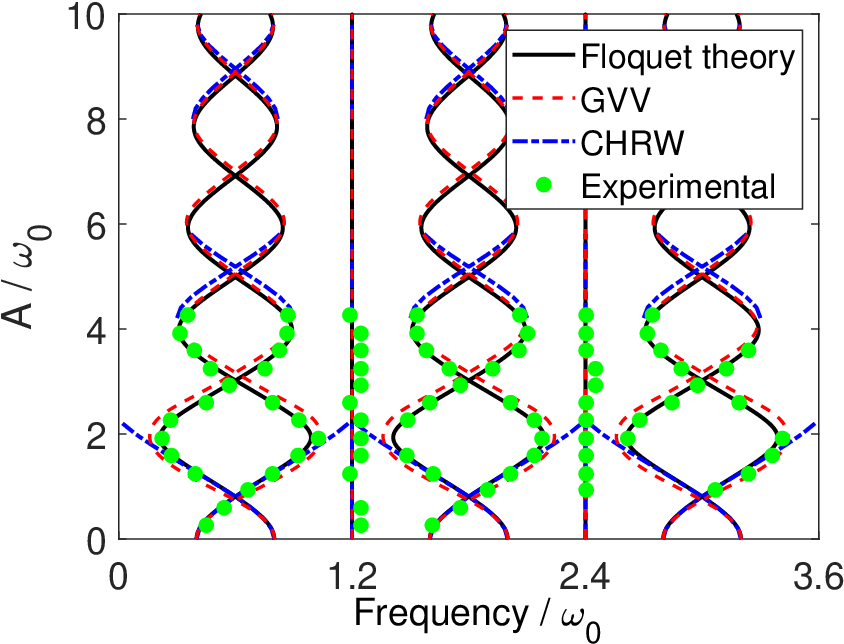}
\caption{\label{fig:dut1}  Frequency components of transition probability $P_1(t)$ as a function of $A$ with $\omega/\omega_0=0.6$. The black lines plot $\pm\Delta\epsilon+2n\omega$ and $2n\omega$, with quasienergy difference $\Delta\epsilon$ determined numerically by Floquet matix (see Appendix~\ref{app:Floquet}). The red dashed lines plot $\pm\Omega+2n\omega$ and $2n\omega$ with $\Omega$ in Eq.~(\ref{eq:our}). The blue dash-dot lines plot $\pm\tilde{\Omega}+2n\omega$ and $2n\omega$ with $\tilde{\Omega}$ in Eq.~(\ref{eq:effrabi}). The green dots are the experimental results extracted from Ref.~\cite{PhysRevLett.115.133601}. }
\end{figure}

\begin{figure}[thp]
\includegraphics[width=3.2in]{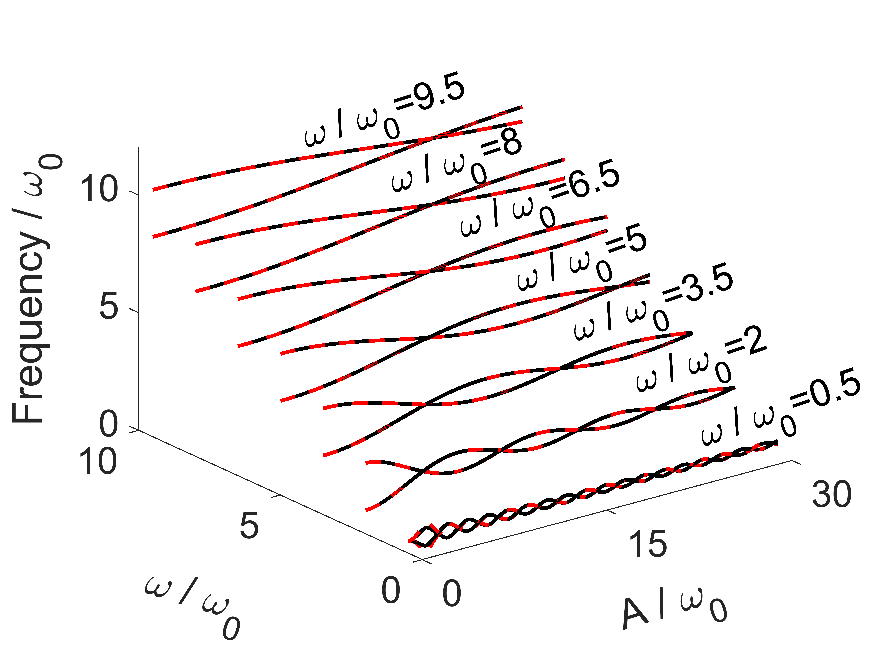}
\caption{\label{fig:dut2} Frequency components of transition probability $P_1(t)$ as a function of $A$ and $\omega$. Same as Fig.~\ref{fig:dut1}, the black lines are the numerical results obtained by Floquet theory and the red dashed lines are the approximate analytical results $\pm\Omega+2n\omega$ with $\Omega$ in Eq.~(\ref{eq:our}).}
\end{figure}

Diagonalize Eq.~(\ref{eq:gvv}) to obtain the eigenvalues, and the Rabi frequency is the difference between the two eigenvalues:
\begin{equation}\label{eq:our}
\Omega=\sqrt{4\delta_{1'0'}\delta_{0'1'}+(\delta_1'-\delta_0'-\omega)^2+B}
\end{equation}
with $B=2(\delta_1'\!-\!\delta_0'-\omega)J_0\omega_0\!-
\!2(\delta_{1'0'}+\delta_{0'1'})J_1\omega_0+(J_0^2+J_1^2)\omega_0^2$. Then all frequencies of the Floquet dynamics can be written as $\pm\Omega+2n\omega$ and $2n\omega$. In Fig.~\ref{fig:dut1}, we compare the numerical, analytical, and the previous experimental results of  system dynamics frequencies in the general off-resonance situation (i.e., $\omega/\omega_0=0.6$). The numerical results are obtained by Floquet theory. The approximate analytical results are obtained by using CHRW and GVV, and the experimental results extracted from Ref.~\cite{PhysRevLett.115.133601}. One can see that the GVV results are more suitable for the exact numerical results than CHRW, because GVV results include higher-order harmonic terms through perturbation theory, which are ignored in CHRW. When $A\in[2.30,4.20]$ or $A\in[6.10,7.99]$, CHRW has no results as discussed in Sec.~\ref{sec:chrw}. In a contrast, the GVV results agree well with the experimental results, and the slight discrepancy may be compensated by considering higher-order perturbation terms. In addition, the GVV results are not only suitable for resonant and near-resonant situations, but also for large detuning situations, as shown in Fig.~\ref{fig:dut2}. In fact, the GVV results are suitable for almost all parameter regimes, except $A/\omega_0\rightarrow0$ and $\omega/\omega_0\rightarrow0$ simultaneously. Therefore, the GVV results provide a better approximation for the Rabi model with arbitrary strength and detuning.


\section{Dissipative open Rabi model}
\label{sec:orm}
In this section, we further consider the dissipative open Rabi model with population damping rate $\Gamma_{ij}$ from state $|i\rangle$ to  $|j\rangle$ and the dephasing rate $\gamma_{ii}$ of state $|i\rangle$. With the assumption of Markovian noise background, the density matrix for the two-level system with basis $|i\rangle$ and $|j\rangle$ evolves according to the Lindblad master equation~\cite{PhysRevB.104.165414,PhysRevB.101.100301},
\begin{eqnarray}\label{eq:orm1}
\dot{\rho}&\!=\!&-i[H(t),\rho ]\!+\!\frac{\Gamma_{ij}}{2}\mathcal{D}[|j\rangle\langle i|]\rho\!+\!\frac{\Gamma_{ji}}{2}\mathcal{D}[|i\rangle\langle j|]\rho, \nonumber \\
  &+&\gamma_{ii}\mathcal{D}[|i\rangle\langle i|]\rho\!+\!\gamma_{jj}\mathcal{D}[|j\rangle\langle j|]\rho,
\end{eqnarray}%
with $\mathcal{D}[\mathcal{O}]\rho=2\mathcal{O}\rho\mathcal{O}^{\dag}-\mathcal{O}^{\dag}\mathcal{O}\rho-\rho\mathcal{O}^{\dag}\mathcal{O}$. For the Rabi model, in the lab frame, we only consider the population damping rate $\Gamma_{10}$ from the excited state $|1\rangle$ to the ground state $|0\rangle$ and the dephasing rate $\gamma_{11}$ of state $|1\rangle$, i.e., $\Gamma_{01}=\gamma_{00}=0$.

The GVV method transforms the Rabi Hamiltonian with basis $|0\rangle$ and $|1\rangle$ in Eq.~(\ref{eq:rabi}) into an effective analytically solvable Hamiltonian in Eq.~(\ref{eq:gvv}) with basis $|s'_{0,1}(t)\rangle$ through two unitary transformations. The decay and excitation rate between $|s'_{0}(t)\rangle$ and $|s'_{1}(t)\rangle$ can be easily calculated by rewriting the lab frame dissipators (i.e., $\Gamma_{10}$ and $\gamma_{11}$) in the new basis (see Appendix~\ref{app:app3}),
\begin{small}
\begin{eqnarray}\label{eq:orm2}
\!\gamma_{s'_1s'_1}(t)&\!=\!&\sin^2\!\left[\frac{A\sin(\omega t)}{\omega}\right]\!\frac{\Gamma_{10}}{8}\!+\!\cos^4\!\left[\frac{A\sin(\omega t)}{2\omega}\right]\!\gamma_{11},\\
\!\gamma_{s'_0s'_0}(t)&\!=\!&\sin^2\!\left[\frac{A\sin(\omega t)}{\omega}\right]\!\frac{\Gamma_{10}}{8}\!+\!\sin^4\!\left[\frac{A\sin(\omega t)}{2\omega}\right]\!\gamma_{11},\\
\!\Gamma_{s'_1s'_0}(t)&\!=\!&\sin^2\!\left[\frac{A\sin(\omega t)}{\omega}\right]\!\frac{\gamma_{11}}{2}\!+\!\cos^4\!\left[\frac{A\sin(\omega t)}{2\omega}\right]\!\Gamma_{10},\\
\!\Gamma_{s'_0s'_1}(t)&\!=\!&\sin^2\!\left[\frac{A\sin(\omega t)}{\omega}\right]\!\frac{\gamma_{11}}{2}\!+\!\sin^4\!\left[\frac{A\sin(\omega t)}{2\omega}\right]\!\Gamma_{10}\label{eq:orm3}.
\end{eqnarray}
\end{small}
Equations~(\ref{eq:orm2})-(\ref{eq:orm3}) show that the decay parameters are time-dependent. One finds that when $A$ is weak (i.e., $A/\omega\rightarrow 0$), $\gamma_{s'_1s'_1}(t)\approx\gamma_{11}$, $\gamma_{s'_0s'_0}(t)\approx 0$, $\Gamma_{s'_1s'_0}(t)\approx\Gamma_{10}$ and $\Gamma_{s'_0s'_1}(t)\approx 0$, because under this condition $|s'_0(t)\rangle\simeq|0\rangle$ and $|s'_1(t)\rangle\simeq|1\rangle$. When $A$ is deep-strong (i.e., $A/\omega\gg1$) , the decay rates  change greatly and periodically with time, and sometimes the excitation rate between $|s'_0\rangle$ and $|s'_1\rangle$ is larger than the decay rate between them, i.e., $\Gamma_{s'_0s'_1}>\Gamma_{s'_1s'_0}$. Similarly, the dephasing rate of $|s'_0\rangle$ can be larger than that of $|s'_1\rangle$, i.e., $\gamma_{s'_0s'_0}>\gamma_{s'_1s'_1}$. These tunable decay parameters provide a new platform for quantum computing~\cite{PhysRevLett.119.150502}.

\begin{figure}[thp]
\includegraphics[width=3.2in]{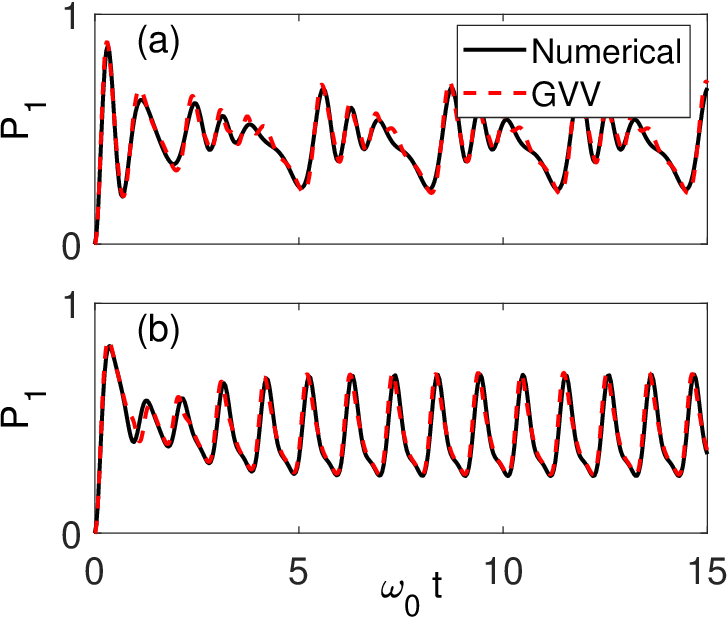}
\caption{\label{fig:orm1} Dynamics of transition probability $P_1$ with $\omega/\omega_0=1$, $A/\omega_0=10$ (a), and $\omega/\omega_0=3$, $A/\omega_0=10$ (b). The black lines denote the numerical results and the red dashed lines are the results obtained by GVV. Here the initial state is $\Psi(0)=|0\rangle$ and decay parameters are $\Gamma_{10}/\omega=1$, $\gamma_{11}/\omega=0.2$ and $\Gamma_{01}/\omega=\gamma_{00}/\omega=0$.}
\end{figure}

Submitting the decay parameters in Eqs.~(\ref{eq:orm2})-(\ref{eq:orm3}) and the Hamiltonian in Eq.~(\ref{eq:gvv}) into the Lindblad master equation in Eq.~(\ref{eq:orm1}), we obtain the dynamics of the transition probability $P_1(t)$, as the red-dashed lines shown in Fig.~\ref{fig:orm1}. To verify the effectiveness of the GVV method, we also present the exact numerical results obtained by solving the Floquet-Lindblad equation with Hamiltonian in Eq.~(\ref{eq:rabi}) and decay parameters $\Gamma_{10}/\omega=1$, $\gamma_{11}/\omega=0.2$, $\Gamma_{01}/\omega=\gamma_{00}/\omega=0$. We find that the GVV results agree well with the numerical results in the strong driving regime, and the small deviations between them may result from higher-order terms of perturbation theory and the loss of the fast oscillation terms when rewriting the dissipation rates in Eqs.~(\ref{eq:orm2})-(\ref{eq:orm3}).

\section{Discussion and conclusion}
\label{sec:con}
In conclusion, by combining DUT with GVV perturbation theories, we present an approximate analytical result of the Rabi model for almost all parameter regimes, except $A/\omega_0\rightarrow0$ and $\omega/\omega_0\rightarrow0$ simultaneously. The GVV results agree well with the numerical solutions obtained by Floquet theory and previous experimental results, and are also beyond the results obtained by using CHRW, which have no results within some parameter ranges. Moreover, as the driving intensity increases from weak to deep-strong, the main frequency component of the Floquet dynamics transitions from Rabi frequency to $2n\omega$. In addition, we further consider the dissipative open Rabi model with population damping and dephasing rates, and the results obtained by GVV Hamiltonian with the regrouping decay rates agree well with the numerical results in the strong driving regime. Our results provide a desired theoretical method for studying the strongly driven closed and open two-level systems. In particular, the development of ultrastrong laser opens the doorway for light-matter interactions in the strong and deep-strong coupling regimes.

The approximate analytical results of the Rabi model are helpful for the use of strong driving for fast quantum gates, specifically qubit state preparation. Starting with the qubit in its ground state, we apply a monochromatic field with driving strength $A$ and frequency $\omega$. With the help of our analytical results, the field strength or duration can be quantitatively calculated to prepare the target state. In addition, the tunable decay rates in the rotating frame may have applications in quantum computing. In particular, the excitation rate between two states can be larger than the decay rate between them, which is almost impossible in the lab frame.

\begin{acknowledgements}
We thank Jianwen Jie for many helpful and intriguing discussions. This work is supported by the National Natural Science Foundation of China under Grants Nos.~12205199 and 12274331, and Natural Science Foundation of Top Talent of SZTU(Grant No. GDRC202202 and GDRC202312).
\end{acknowledgements}

\appendix
\section{Derivation of Eq.~(\ref{eq:chrwp2})}\label{app:chrw}
The wave function in basis $|s_{1,0}(t)\rangle$ is $|\Psi(t)\rangle=c_1(t)|s_1(t)\rangle+c_0(t)|s_0(t)\rangle$. The initial condition after the unitary transformation is invariant, $|s_{i}(0)\rangle=|i\rangle, i=1,0$. For an initial state in $|0\rangle$, we can solve the Schr$\ddot{\text{o}}$dinger equation with Hamiltonian in Eq.~(\ref{eq:u1h12}) easily as given in Ref.~\cite{QO,PhysRevA.86.023831},
\begin{eqnarray}\label{eq:solu1}
c_1(t)&=&-\frac{i\tilde{A}}{2\tilde{\Omega}}\sin\left(\frac{\tilde{\Omega} t}{2}\right)e^{i\frac{\omega t}{2}}\\
c_0(t)&=&\left[\cos\left(\frac{\tilde{\Omega} t}{2}\right)\!+\!\frac{i\tilde{\Delta}}{\tilde{\Omega}}\sin\left(\frac{\tilde{\Omega} t}{2}\right)\right]e^{-i\frac{\omega t}{2}}.\label{eq:solu0}
\end{eqnarray}
The population of excited state $|1\rangle$ can be expressed as
\begin{equation}\label{eq:chrwp1}
P_1(t)\!=\!\left|\cos\left[\frac{A\xi\sin(\omega t)}{2\omega}\right]c_1(t)\!-\!i\sin\left[\frac{A\xi\sin(\omega t)}{2\omega}\right]c_0(t)\right|^2
\end{equation}
Substituting Eq.~(\ref{eq:solu1}) and Eq.~(\ref{eq:solu0}) into Eq.~(\ref{eq:chrwp1}), and regrouping terms with the same frequency, we have
\begin{eqnarray}\label{eq:chrwpp}
P_1(t)&=&C_0+C_1\cos(\tilde{\Omega} t)+\sum_{n=1}^{\infty}[C_{2_n}\cos(2n\omega t+\tilde{\Omega} t)\nonumber \\
&+&C_{3_n}\cos(2n\omega t-\tilde{\Omega} t)+C_{4_n}\cos(2 n\omega t)],
\end{eqnarray}
with
\begin{eqnarray}\label{eq:c0}
C_0&=&\frac{1}{2}-\frac{\tilde{\Delta}^2}{2\tilde{\Omega}^2}J_0\left(\frac{A\xi}{\omega}\right)+\frac{\tilde{\Delta}\tilde{A}}{4\tilde{\Omega}^2}
J_1\left(\frac{A\xi}{\omega}\right),\\
C_1&=&-\frac{\tilde{A}^2}{8\tilde{\Omega}^2}J_0\left(\frac{A\xi}{\omega}\right)-\frac{\tilde{\Delta}\tilde{A}}{4\tilde{\Omega}^2}
J_1\left(\frac{A\xi}{\omega}\right),\\
C_{2_n}&=&-\left[\frac{\tilde{A}^2}{8\tilde{\Omega}^2}+\frac{n\omega\tilde{A}}{2A\xi\tilde{\Omega}}\right]
J_{2n}\left(\frac{A\xi}{\omega}\right)\nonumber \\
&+&\frac{\tilde{\Delta}\tilde{A}}{8\tilde{\Omega}^2}\left[J_{2n-1}\left(\frac{A\xi}{\omega}\right)-
J_{2n+1}\left(\frac{A\xi}{\omega}\right)\right],\\
C_{3_n}&=&-\left[\frac{\tilde{A}^2}{8\tilde{\Omega}^2}-\frac{n\omega\tilde{A}}{2A\xi\tilde{\Omega}}\right]
J_{2n}\left(\frac{A\xi}{\omega}\right)\nonumber \\
&+&\frac{\tilde{\Delta}\tilde{A}}{8\tilde{\Omega}^2}\left[J_{2n-1}\left(\frac{A\xi}{\omega}\right)-
J_{2n+1}\left(\frac{A\xi}{\omega}\right)\right],
\end{eqnarray}
\begin{eqnarray}\label{eq:c4}
C_{4_n}&=&\frac{\tilde{\Delta}\tilde{A}}{4\tilde{\Omega}^2}
\left[J_{2n+1}\left(\frac{A\xi}{\omega}\right)-J_{2n-1}\left(\frac{A\xi}{\omega}\right)\right]\nonumber\\
&-&\frac{\tilde{\Delta}^2}{\tilde{\Omega}^2}J_{2n}\left(\frac{A\xi}{\omega}\right),
\end{eqnarray}
where we used the formula $\cos[A\sin(\omega t)/\omega]=\sum_{n=-\infty}^{\infty}J_n(A/\omega)\cos(n\omega t)$.

\section{Derivation of the $2\times2$ effective Hamiltonian by the GVV theory}\label{app:app1}
In this section, we reduce the infinite-dimensional Floquet matrix in Eq.~(\ref{eq:dut7}) into a $2\times2$ effective matrix by using GVV perturbation theory~\cite{PhysRevA.32.377,PhysRevA.101.022108}. Consider the Floquet states $|s'_1,0\rangle$ nearly degenerate with $|s'_0,1\rangle$. According to the perturbation theory, we expand the $2\times2$ matrix $h$ and its eigenstates $\Phi$ in powers of $\omega_0$. The zeroth and higher-order of $\Phi$ are given by
\begin{small}
\begin{eqnarray}\label{eq:phi0}
\Phi_1'^{(0)}&\!=\!&|s_1',0\rangle,~~
\Phi_0^{(0)}=|s_0',1\rangle, \nonumber\\
\Phi_1'^{(1)}&\!=\!&\sum_{\begin{subarray}{c} k=-\infty \\ k\neq 0\end{subarray}}^{\infty} \left[\frac{-J_{2k+1}\omega_0/2}{J_0\omega_0\!-\!(2k+1)\omega}|s_0',2k\!+\!1\rangle\!-\!\frac{J_{2k}\omega_0}{4k\omega}|s_1',2k\rangle\right],\nonumber \\
\Phi_0^{(1)}&\!=\!&\sum_{\begin{subarray}{c} k=-\infty \\ k\neq 0\end{subarray}}^{\infty} \left[\frac{-J_{2k-1}\omega_0/2}{J_0\omega_0\!+\!(2k-1)\omega}|s_1',2k\rangle\!+\!\frac{J_{2k}\omega_0}{4k\omega}|s_0',2k\!+\!1\rangle\right].\nonumber\\
\end{eqnarray}
\end{small}
The zeroth and higher-order of $h$ represented by $\Phi$ are
\begin{equation}\label{eq:h0}
 h^{(0)}=\left(
 \begin{array}{ c c}
 \frac{J_0\omega_0}{2}&0\\
 0&\omega-\frac{J_0\omega_0}{2}
 \end{array}
 \right),
\end{equation}
\begin{equation}\label{eq:h1}
 h^{(1)}=\langle\Phi^{(0)}|V'|\Phi^{(0)}\rangle
=\left(
 \begin{array}{ c c}
 0&-\frac{J_1\omega_0}{2}\\
 -\frac{J_1\omega_0}{2}&0
 \end{array}
 \right),
\end{equation}
\begin{eqnarray}\label{eq:h2}
h^{(2)}&=&\langle\Phi^{(0)}|V'|\Phi^{(1)}\rangle-h^{(1)}\langle\Phi^{(0)}|\Phi^{(1)}\rangle \nonumber \\
&=&\left(
\begin{array}{ c c}
 \delta_{1'}&\delta_{1'0'}\\
 \delta_{0'1'}&\delta_{0'}
 \end{array}
 \right),
\end{eqnarray}
with $\delta_{1'}$, $\delta_{0'}$, $\delta_{1'0'}$ and $\delta_{0'1'}$ given in Eqs.~(\ref{eq:delta1})-(\ref{eq:delta4}).
Equations (\ref{eq:h0}) and (\ref{eq:h1}) form a first-order perturbation matrix:
\begin{equation}\label{eq:grwa}
 H_{\text{GRWA}}=\left(
 \begin{array}{ c c}
 \frac{J_0\omega_0}{2}&-\frac{J_1\omega_0}{2}\vspace{2mm}\\
 -\frac{J_1\omega_0}{2}&\omega-\frac{J_0\omega_0}{2}
 \end{array}
 \right),
\end{equation}
which is similar to the rotating wave approximation and neglects all other non-resonant coupling terms. Note that this RWA is different from the conventional one subject to the transverse coupling, where RWA breaks down in the strong field, and we refer it to generalized rotating wave approximation(GRWA). Diagonalize Eq.~(\ref{eq:grwa}) to obtain the eigenvalues, and the Rabi frequency is the difference between the two eigenvalues:
\begin{equation}\label{eq:ourgrwa}
\Omega'=\sqrt{(\omega-J_0\omega_0)^2+J_1^2\omega_0^2}.
\end{equation}
Then all frequencies of the Floquet dynamics can be written as $\pm\Omega'+2n\omega$ and $2n\omega$.

\begin{figure}[thp]
\includegraphics[width=3.2in]{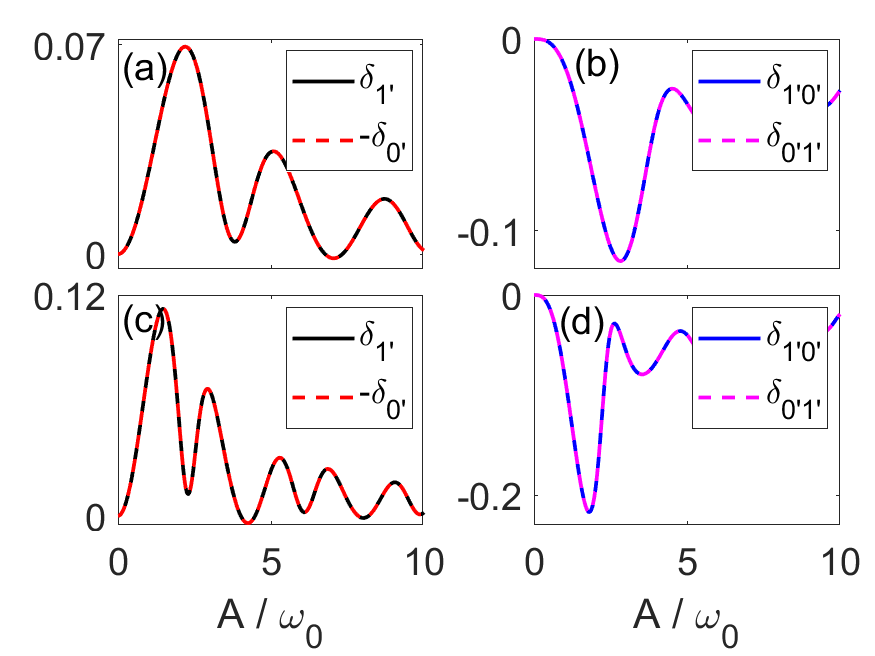}
\caption{\label{fig:app1}  Comparison of $\delta_{1'}$ and $\delta_{0'}$, $\delta_{1'0'}$ and $\delta_{0'1'}$ for various $A$ with $\omega/\omega_0=1$(a),(b) and $\omega/\omega_0=0.6$(c),(d).}
\end{figure}

\begin{figure}[thp]
\includegraphics[width=3.2in]{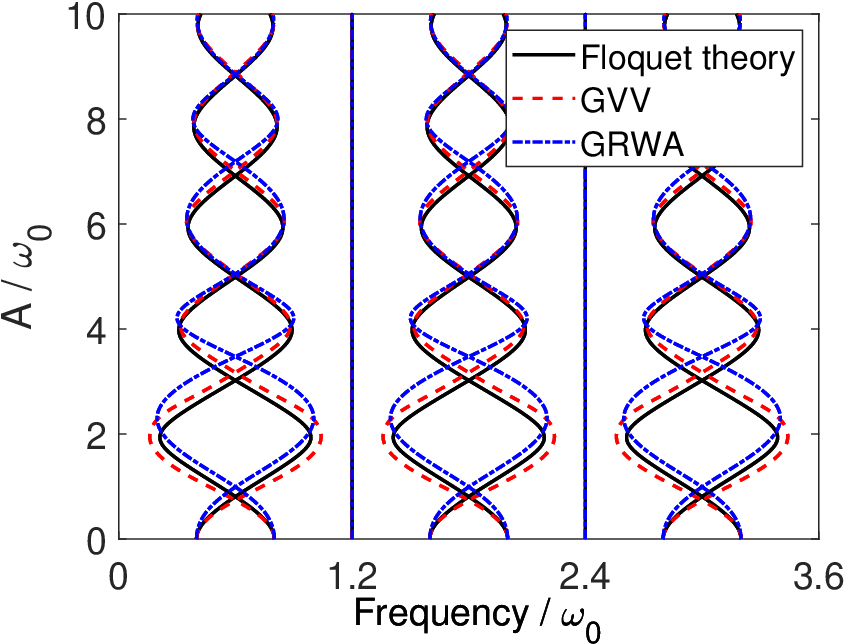}
\caption{\label{fig:app2}  Comparison of numerical results obtained by Floquet theory and approximate analytical GRWA results [i.e., $\pm\Omega'+2n\omega$ and $2n\omega$ with $\Omega$ in Eq.~(\ref{eq:ourgrwa})] and GVV results [i.e., $\pm\Omega+2n\omega$ and $2n\omega$ with $\Omega$ in Eq.~(\ref{eq:our})] results of the frequency components for various $A$ with $\omega/\omega_0=0.6$.}
\end{figure}

Equations (\ref{eq:h0})-(\ref{eq:h2}) form the second-order perturbation matrix shown in Eq.~(\ref{eq:gvv}). In Fig.~\ref{fig:app1}, we plot $\delta_{1'}$, $\delta_{0'}$, $\delta_{1'0'}$ and $\delta_{0'1'}$ as a function of $A$ for resonant and off-resonant situations. We find that $\delta_{0'}=-\delta_{1'}$, $\delta_{1'0'}=\delta_{0'1'}$ with $\omega/\omega_0=1$ and $\omega/\omega_0=0.6$. In Fig.~\ref{fig:app2}, we compare the numerical and analytic results of the frequencies obtained by GVV Hamiltonian and GRWA Hamiltonian. Clearly, the GVV results fit better than the GRWA to the exact numerical results, indicating the deviation of the GRWA and the validity of the GVV.

\section{Numerical results obtained by Floquet theory}\label{app:Floquet}

According to the Floquet theory, the Rabi model in Eq.~(\ref{eq:rabi}) can be rewritten as 
\begin{equation}\label{eq:app21}
H=H^{[0]}+H^{[-1]}e^{i\omega t}+H^{[1]}e^{-i\omega t}
\end{equation}
with $H^{[0]}=\omega_0\sigma_z/2$ and $H^{[\pm1]}=A\sigma_x/4$. Based on Eq.~(\ref{eq:fme}), the Floquet matrix is
\begin{widetext}
\begin{equation}
\label{eq:app22}
\renewcommand\arraystretch{1.7}
H''_F=\left(
\begin{array}{c c c|c c|c c|c c|c c c}
\ddots&&&&&&&&&&&\iddots\\
 \hline
   &\frac{\omega_0}{2}-2\omega & 0&0& \frac{A}{4}&0&0&0&0&0&0&\\
   &0&\frac{-\omega_0}{2}-2\omega&\frac{A}{4}&0&0&0&0&0&0&0&\\
   \hline
  &0&\frac{A}{4}&\frac{\omega_0}{2}-\omega &0&0&\frac{A}{4}&0&0&0&0&\\
  &\frac{A}{4}&0&0&\frac{-\omega_0}{2}-\omega&\frac{A}{4}&0&0&0&0&0&\\
  \hline
  &0&0&0 & \frac{A}{4}& \frac{\omega_0}{2}&0& 0&\frac{A}{4}&0&0&\\
  &0&0&\frac{A}{4}&0&0&\frac{-\omega_0}{2}&\frac{A}{4}&0&0&0&\\
  \hline
  &0&0&0&0&0&\frac{A}{4} & \frac{\omega_0}{2}+\omega &0&0&\frac{A}{4}&\\
  &0&0&0&0&\frac{A}{4}&0&0&\frac{-\omega_0}{2}+\omega&\frac{A}{4}&0&\\
  \hline
  &0&0&0&0&0&0&0&\frac{A}{4}&\frac{\omega_0}{2}+2\omega&0&\\
  &0&0&0&0&0&0&\frac{A}{4}&0&0&\frac{-\omega_0}{2}+2\omega&\\
   \hline
\iddots&&&&&&&&&&&\ddots
\end{array}
\right).
\renewcommand\arraystretch{1.7}
\begin{array}{c c}
\\
\leftarrow&|1,-2\rangle\\
\leftarrow&|0,-2\rangle\\
\leftarrow&|1,-1\rangle\\
\leftarrow&|0,-1\rangle\\
\leftarrow&|1,0\rangle\\
\leftarrow&|0,0\rangle\\
\leftarrow&|1,+1\rangle\\
\leftarrow&|0,+1\rangle\\
\leftarrow&|1,+2\rangle\\
\leftarrow&|0,+2\rangle\\\nonumber
\end{array}
\end{equation}
\end{widetext}
We obtain quasienergies by numerically diagonalize $H''_F$. In our calculations, we truncate the matrix $H''_F$ with n ranging from -30 to 30.

\section{Unitary transformation of decay parameters}\label{app:app3}
In this section, we give the derivation of Eqs.~(\ref{eq:orm2})-(\ref{eq:orm3}). In the lab frame, the master equation of the dissipative open Rabi model with population damping rate $\Gamma_{10}$ and dephasing rate $\gamma_{11}$ is~\cite{PhysRevA.105.063724}
\begin{equation}\label{eq:lmetwo}
\dot{\rho}=-i[H(t),\rho ]+\frac{\Gamma_{10}}{2}\mathcal{D}[\sigma_{01}]\rho+\gamma_{11}\mathcal{D}[\sigma_{11}]\rho.
\end{equation}
According to Eq.~(\ref{eq:dut8}), the rotation matrix is
\begin{equation}
U_4(t)=\cos\left[ \frac{A\sin(\omega t)}{2\omega}\right]I+i\sin\left[ \frac{A\sin(\omega t)}{2\omega}\right]\sigma_x,
\end{equation}
with $I$ a $2\times2$ identity matrix. The decay and excitation rate between $|s'_1(t)\rangle$ and $|s'_0(t)\rangle$ can be easily calculated by rewriting the lab frame dissipators in the new basis~\cite{PhysRevLett.119.150502},
\begin{eqnarray}\label{eq:decay}
\frac{\Gamma_{10}}{2}\mathcal{D}[\sigma_{01}]\rho&=&\frac{\Gamma_{10}}{2}\mathcal{D}[U_4^{\dagger}(t)\sigma_{s'_0s'_1}U_4(t)]\rho\nonumber \\
&=&\frac{\Gamma_{10}}{2}\mathcal{D}\left[\zeta(\sigma_{s'_0s'_0}\!-\!\sigma_{s'_1s'_1})\!+\!\beta\sigma_{s'_1s'_0}\!+\!\eta\sigma_{s'_0s'_1}\right]\rho, \nonumber\\
\\
\gamma_{11}\mathcal{D}[\sigma_{11}]\rho&=&\gamma_{11}\mathcal{D}[U_4^{\dagger}(t)\sigma_{s'_1s'_1}U_4(t)]\rho\nonumber \\
&=&\gamma_{11}\mathcal{D}\left[\beta\sigma_{s'_0s'_0}\!+\!\eta\sigma_{s'_1s'_1}\!+\!\zeta(\sigma_{s'_1s'_0}\!-\!\sigma_{s'_0s'_1})\right]\rho\nonumber \\
\end{eqnarray}
with $\zeta=\frac{1}{2}\sin\left[ \frac{A\xi\sin(\omega t)}{2\omega}\right]$, $\beta=\sin^2\left[ \frac{A\xi\sin(\omega t)}{2\omega}\right]$ and $\eta=\cos^2\left[ \frac{A\xi\sin(\omega t)}{2\omega}\right]$.  Therefore, by regrouping the above dissipators and dropping out the fast oscillating terms, such as $\sigma_{s'_0s'_1}\rho\sigma_{s'_0s'_1}$ and $\sigma_{s'_1s'_0}\rho\sigma_{s'_1s'_0}$ etc., we obtain the effective decay rates in Eqs.~(\ref{eq:orm2})-(\ref{eq:orm3}).

%\bibliography{ref}

\end{document}